\documentclass{Interspeech2024}

\usepackage{amsmath}
\usepackage{nccmath}
\usepackage{cite}
\usepackage{amssymb}
\newcommand{\blskip}{\baselineskip   10.14pt}
\newcommand{\ftblskip}{\baselineskip 10.14pt}

\def\X{{\mathbf{X}}}

\def\0{{\mathbf{0}}}

\def\Pe2e{P_{\mathtt{e2e}}}

\def\w{{\mathbf{w}}}
\def\Pasr{P_{\mathtt{asr}}}
\def\Plm{P_{\mathtt{lm}}}

\def\best{{\mathtt{best}}}

\interspeechcameraready

\title{Applying LLMs for rescoring N-best ASR hypotheses of casual conversations:\\Effects of domain adaptation and context carry-over}

\name{Atsunori}{Ogawa}
\name{Naoyuki}{Kamo}
\name{Kohei}{Matsuura}
\name{Takanori}{Ashihara}
\name{Takafumi}{Moriya}
\name{\\Takatomo}{Kano}
\name{Naohiro}{Tawara}
\name{Marc}{Delcroix}

\address{NTT Corporation, Japan}
\email{atsunori.ogawa@ntt.com}


\begin{document}
\blskip 

\maketitle

\begin{abstract}
\vspace{1.50mm}
Large language models (LLMs) have been
successfully
applied for rescoring automatic speech recognition (ASR) hypotheses.
However, their ability to rescore ASR hypotheses
of casual conversations
has not been
sufficiently
explored.
In this study,
we reveal
it
by
performing
$N$-best ASR hypotheses rescoring
using Llama2
on the CHiME-7
distant ASR
(DASR)
task.
Llama2
is one of the most representative LLMs,
and
the CHiME-7
DASR
task
provides datasets of
casual conversations between multiple participants.
We
investigate
the effects of 
domain adaptation of the
LLM
and
context carry-over
when performing $N$-best rescoring.
Experimental results show that,
even without domain adaptation,
Llama2 outperforms a standard-size domain-adapted Transformer-LM,
especially when using a long context.
Domain adaptation shortens the context length needed with Llama2
to achieve its best performance,
i.e., it
reduces
the computational cost of Llama2.
\end{abstract}
%
\vspace{-1.50mm}
\section{Introduction}
\label{sec_intro}
%
Large language models (LLMs),
such as GPT-4 \cite{OpenAI_arXiv2023},
PaLM2 \cite{Google_arXiv2023},
and Llama2
(Large Language Model META AI) \cite{Touvron_arXiv2023a},
have now become a prominent component in modern natural language processing
(NLP)
and are
successfully utilized in various NLP tasks,
such as machine translation, text summarization, and question answering.
Recently,
they have been
used
not only in NLP tasks
but also in speech-related tasks,
including automatic speech recognition (ASR).
%
A simple
way to utilize LLMs in ASR
is using them in the second-pass rescoring (re-ranking)
of multiple ASR hypotheses
represented
as
an $N$-best list or a lattice,
which is obtained by the first-pass ASR decoding.
%
Several studies have reported
the usefulness of
LLMs
in $N$-best or lattice rescoring of ASR hypotheses
\cite{Shin_PMLR2019,Li_ICASSP2020,Salazar_ACL2020,Chiu_SLT2021,Fohr_IS2021,Zheng_ASRU2021,Futami_ASRU2021,Xu_ICASSP2022,Udagawa_IS2022,Chen_ICASSP2023,Yu_ASRU2023,Li_ASRU2023,Shivakumar_ASRU2023}.

Thanks to the significant progress of end-to-end (E2E) neural network modeling,
the performance of ASR has greatly improved.
Despite this
significant
progress,
ASR accuracy remains unsatisfactory in some situations,
such as performing ASR 
in daily-life environments
\cite{Cornell_CHiME2023,Wang_CHiME2023,Ye_CHiME2023,Kamo_CHiME2023,Prisyach_CHiME2023,Park_CHiME2023}.
%
The distant ASR (DASR) task of the CHiME-7 challenge
provides a dataset
of such challenging situations \cite{Cornell_CHiME2023}.
The dataset contains casual conversations between multiple participants
at
real dinner parties.
LMs can be expected to
play an
important
role
in ASR of such casual conversational speech, 
and most of the
submitted
systems
try to use
LMs
during
ASR decoding
and/or for rescoring ASR hypotheses
\cite{Ye_CHiME2023,Kamo_CHiME2023,Prisyach_CHiME2023,Park_CHiME2023}.
However,
the effect of using LMs is limited
(the
first-place system
does not use
any
LMs \cite{Wang_CHiME2023}),
%
and there
is a demand for LMs
to
deal with
such highly
casual conversational speech.

As described above,
several studies have successfully applied LLMs for
rescoring ASR hypotheses
\cite{Shin_PMLR2019,Li_ICASSP2020,Salazar_ACL2020,Chiu_SLT2021,Fohr_IS2021,Zheng_ASRU2021,Futami_ASRU2021,Xu_ICASSP2022,Udagawa_IS2022,Chen_ICASSP2023,Yu_ASRU2023,Li_ASRU2023,Shivakumar_ASRU2023}.
However,
their
targets are not casual conversations,
and the ability of LLMs to rescore ASR hypotheses of
casual conversations
remains unclear
(note that LLMs are not allowed to be used in the CHiME-7
challenge
\cite{Cornell_CHiME2023}).
In this study,
we reveal
it
by
performing
$N$-best ASR hypotheses rescoring
using
Llama2-7B \cite{Touvron_arXiv2023a},
which is one of the most representative
Transformer
\cite{Vaswani_NIPS2017}
decoder-based
causal
LLMs,
on the CHiME-7 DASR task.
We comprehensively investigate
the effects of 
domain adaptation of the
LLM
and
context carry-over
\cite{Zheng_ASRU2021,Udagawa_IS2022,Chen_ICASSP2023,Ye_CHiME2023}
when performing $N$-best rescoring.
We employ QLoRA
\cite{Dettmers_NeurIPS2023}
for
memory efficient
domain adaptation
and consider various context lengths (up to 1024 tokens)
in context carry-over.

We conducted experiments,
including experimental settings that have not been investigated
in previous studies
\cite{Shin_PMLR2019,Li_ICASSP2020,Salazar_ACL2020,Chiu_SLT2021,Fohr_IS2021,Zheng_ASRU2021,Futami_ASRU2021,Xu_ICASSP2022,Udagawa_IS2022,Chen_ICASSP2023,Yu_ASRU2023,Li_ASRU2023,Shivakumar_ASRU2023},
and thus,
the experimental results
and findings
obtained in this study are informative
for researchers in this field
(note that
Llama2-7B
is allowed to be used
in the CHiME-8 challenge \cite{CHiME}).
Our main findings can be summarized as follows.
%
\begin{itemize}
\item Even without domain adaptation,
Llama2
significantly
outperforms a standard-size domain-adapted Transformer-LM.
\item Both
domain adaptation and context carry-over improve the Llama2 performance.
\item Even without domain adaptation,
by considering a very long context (e.g., 1024 tokens),
Llama2 captures the flow of a conversation
and achieves
the lowest word error rate (WER),
which is
achieved
with the domain-adapted Llama2.
\item Domain adaptation
shortens the context length
needed with Llama2 to achieve the lowest WER,
significantly reducing
the computational cost of Llama2.
%
%
\end{itemize}
%
\vspace{-3.00mm}
\section{Relation to prior work}
\label{sec_prior}
\vspace{-0.50mm}
%
Previous studies
\cite{Shin_PMLR2019,Li_ICASSP2020,Salazar_ACL2020,Chiu_SLT2021,Fohr_IS2021,Zheng_ASRU2021,Futami_ASRU2021,Xu_ICASSP2022,Udagawa_IS2022,Chen_ICASSP2023,Yu_ASRU2023,Li_ASRU2023,Shivakumar_ASRU2023}
use both Transformer encoder-based
bidirectional
LLMs,
such as
BERT
\cite{Devlin_NAACLHLT2019},
RoBERTa
\cite{Liu_arXiv2019},
and
ELECTRA
\cite{Clark_ICLR2020},
and Transformer decoder-based
unidirectional
LLMs,
such as
GPT
\cite{Radford_OpenAITechRep2018},
GPT-2
\cite{Radford_OpenAITechRep2019},
PaLM
\cite{Chowdhery_arXiv2022}
and
Llama1
\cite{Touvron_arXiv2023b},
but focus more on the former encoder-based LLMs.
In contrast,
in this study,
we
focus on
a decoder-based LLM,
i.e., Llama2 \cite{Touvron_arXiv2023a},
since recently released LLMs are mainly decoder-based,
e.g.,
GPT-4 \cite{OpenAI_arXiv2023},
PaLM2 \cite{Google_arXiv2023},
and Llama2,
and we can expect their further progress.

%
Some
previous studies
\cite{Li_ICASSP2020,Chiu_SLT2021,Zheng_ASRU2021,Xu_ICASSP2022,Udagawa_IS2022,Yu_ASRU2023}
use
moderately
conversational datasets,
such as
Switchboard (conversations on telephone calls)
\cite{Godfrey_ICASSP1992},
AMI (conversations on meetings)
\cite{Carletta_MLMI2005},
and
an in-house dataset
(conversations with a conversational agent)
\cite{Xu_ICASSP2022,Yu_ASRU2023}.
In contrast,
in this study,
we use the CHiME-7 DASR task
dataset
(conversations at dinner parties)
\cite{Cornell_CHiME2023},
which is much more 
casual and challenging
than the above datasets,
to reveal the applicability of LLMs for rescoring
ASR hypotheses of highly casual conversations.

Considering
past and future contexts is
useful
for rescoring current ASR hypotheses,
and some previous studies
perform context carry-over
\cite{Zheng_ASRU2021,Udagawa_IS2022,Chen_ICASSP2023}.
The past context is used with both
encoder-based bidirectional LLMs
and
decoder-based unidirectional LLMs,
while
the future context is used only with encoder-based LLMs.
In this study,
we utilize only the past context since we use Llama2,
but we comprehensively investigate the effect
of the context length by varying it in a wide range,
i.e., 0 (without considering the context) to 1024 tokens.
The context length investigated in this study is much longer
than that investigated in the previous studies,
i.e., up to 180 tokens \cite{Zheng_ASRU2021}.
%
\section{Models and methods}
\label{sec_mandm}
\vspace{-0.00mm}
%
We introduce the LMs used in this study,
the domain adaptation methods of the LMs,
the $N$-best rescoring method with context-carry over,
and text preprocessing.
%
\vspace{-1.30mm}
\subsection{Language models}
\label{ssec_lms}
\vspace{-0.35mm}
%
We use Llama2-7B
\cite{Touvron_arXiv2023a}
as the main LLM.
As a competitor,
we also prepared a standard-size Transformer-LM.
%
We used the Llama2 tokenizer
(its
vocabulary size is 32k BPE
\cite{Sennrich_ACL2016,Kudo_EMNLP2018}
tokens)
as that of the standard-size Transformer-LM,
and thus,
we can fairly compare these two models in
terms of
perplexity (PPL).
%
To build
the standard-size Transformer-LM,
we first copied the configuration of Llama2-7B
and edited it to define a downsized model structure,
and then
we trained the configurated model from scratch using a text
dataset.
The model size (number of model parameters) is
about
70M,
i.e.,
1/100 of
the Llama2-7B size,
which
is
the standard size of a Transformer-LM.
This model inherits
the
configuration
of
Llama2-7B,
and thus,
in this study,
we refer to it as Slama2-70M,
i.e., Standard-size (or Smaller-size) of Llama2.
%
%
Details of Slama2-70M
are
described in Section~\ref{ssec_sets}.

We also use Llama2-7B-Chat,
which is a
fine-tuned
version of Llama2-7B
that is optimized for dialogue use cases
\cite{Touvron_arXiv2023a},
since it may be more suitable than the base Llama2-7B
for rescoring ASR hypotheses of casual conversation.
We investigate which model is more suitable for the target
in Section~\ref{ssec_ppls}.
%
\vspace{-1.30mm}
\subsection{Domain adaptation}
\label{ssec_adapt}
\vspace{-0.35mm}
%
Llama2 is trained using massive text
datasets
and is expected to 
have general linguistic knowledge.
However,
conversations contained in the CHiME-7 DASR task
dataset
are highly casual,
and thus,
transcriptions of such conversations may not be included in the
Llama2 training text
datasets
(their details are not opened \cite{Touvron_arXiv2023a}).
We employ QLoRA
\cite{Dettmers_NeurIPS2023}
to adapt
Llama2
to the target casual conversational domain
with its memory efficient way.
With QLoRA,
a 4-bit quantized large number of the LLM parameters are frozen,
while a small number of low-rank adapters (LoRA)
\cite{Hu_ICLR2022}
are fine-tuned
using a smaller-size target-domain text dataset.
As regards domain adaptation of
Slama2,
we perform full parameter fine-tuning.
Details of domain adaptation are described in Section~\ref{ssec_sets}.
%
\vspace{-1.30mm}
\subsection{N-best rescoring with context carry-over}
\label{ssec_nbest}
\vspace{-0.35mm}
%
Let $\X_{i}$ be a feature vector sequence of the $i$th utterance
in an input utterance sequence.
%
As the first-pass ASR
decoding,
an E2E ASR model decodes $\X_{i}$
and outputs $N$-best ASR hypotheses
(an $N$-best list)
of the input utterance as
$\{\w_{i}^{r}\}_{r=1}^{N}$,
where $\w_{i}^{r}$ is the $r$th rank hypothesis (token sequence).
The ASR model provides the score (log-probability)
for each of the $N$-best hypotheses
as $\{\log{\Pasr}(\w_{i}^{r}|\X_{i})\}_{r=1}^{N}$.

Then,
as the second-pass
post-processing,
we perform $N$-best rescoring.
We first
calculate
the
LM score (log-probability)
for each of the $N$-best hypotheses
as $\{\log{\Plm}(\w_{i}^{r})\}_{r=1}^{N}$
using an LM.
%
Next,
for each rank,
i.e., $r=1,{\cdots},N$,
we combine the ASR and LM scores as,
%
%
%
\begin{equation}
\label{eq_nbest}
{\log}{P}(\w_{i}^{r}|\X_{i})
= \log{\Pasr}(\w_{i}^{r}|\X_{i})
+ \alpha\log{\Plm}(\w_{i}^{r})
+ \gamma\lvert{\w_{i}^{r}}\rvert,
\end{equation}
where
$\alpha$ ($\alpha \geq 0$) is the language weight
and $\gamma$ ($\gamma \geq 0$) is the reward
that is given proportional to the length of $\w_{i}^{r}$.
%
%
Lastly,
we select the best
(the highest score rank)
hypothesis
based on the combined score
${\log}P(\w_{i}^{r}|\X_{i})$
in Eq.~(\ref{eq_nbest})
as the final 1-best ASR
hypothesis.
%
%
%

%
In the above basic $N$-best rescoring procedure,
we focus on the current hypotheses.
%
However,
considering
the past hypotheses sequence
as the context
is effective
for rescoring the current hypotheses,
especially for the conversational speech case.
In this study,
as with some previous studies
\cite{Zheng_ASRU2021,Udagawa_IS2022,Chen_ICASSP2023,Ye_CHiME2023},
we perform context carry-over in $N$-best rescoring.
%
To consider the context,
we modify the LM score
in Eq.~(\ref{eq_nbest})
as,
\begin{equation}
\label{eq_cco}
\log{\Plm}(\w_{i}^{r}) \rightarrow \log{\Plm}(\w_{i}^{r}|\w_{-L:-1}^{\best}),
\end{equation}
where $\w_{-L:-1}^{\best}$ is the best past context (token sequence)
of the length (number of tokens) $L$
obtained by $N$-best rescoring for the past $N$-best hypotheses sequence.
Note that,
in this study,
we do not care about the hypothesis (utterance) boundaries,
i.e., the best past context can start
from the middle of a past 1-best hypothesis.
Note also that,
as with $N$-best rescoring,
we can perform
PPL
calculation
with context-carry over.
%
We
comprehensively investigate the effect
of the context length
$L$
by varying it in a wide range
in Section~\ref{ssec_wers}.
%
\vspace{-1.80mm}
\subsection{Text processing}
\label{ssec_text}
\vspace{-0.85mm}
%
The authors of \cite{Ye_CHiME2023},
who submitted
the second-place
system of the CHiME-7
challenge,
ordered
utterances (sentences) in the training text
dataset
as,
speaker 1's utterance 1, utterance 2, ...,
speaker 2's utterance 1, utterance 2, ...,
and trained an LM
(they performed $N$-best rescoring
by applying the same ordering to ASR hypotheses).
%
This speaker-conditioned ordering is based on the assumption
that utterances from
one
speaker have some consistency,
and,
within the speaker,
the past utterances are
useful
in predicting the current utterance.
However,
this ordering ignores the flow of a conversation.
We investigate which of the speaker-conditioned order
or the conversational order is more suitable for the CHiME-7 DASR task
in Section~\ref{ssec_ppls}.

Llama2 is trained using texts
that preserve
their original forms
\cite{Touvron_arXiv2023b,Touvron_arXiv2023a},
i.e.,
the texts preserve
capitalized characters
and symbols,
such as
commas,
periods,
(double) quotations,
\mbox{(semi-)} colons,
question/exclamation marks,
and so on.
%
In contrast,
texts
used in the ASR research field,
including texts in the CHiME-7 DASR task dataset,
are usually heavily normalized,
i.e.,
all the characters in the texts are lowercased,
and all the symbols are removed from the texts.
It is not clear whether Llama2 can appropriately treat
these heavily normalized texts.
However,
what
we can do to recover the original texts is limited.
%
In this study,
we add a period
for each sentence (or hypothesis in $N$-best rescoring).
What else we can do is capitalize the first character for each sentence
(but it is difficult to recover other capitalization, e.g., named entities).
We investigate whether this capitalization of the first character is
effective
for
Llama2 in Section~\ref{ssec_ppls}.
%
\vspace{-4.00mm}
\section{Experiments}
\label{sec_exps}
\vspace{-0.00mm}
%
We conducted $N$-best rescoring experiments
using the CHiME-7 DASR task dataset
\cite{Cornell_CHiME2023}
on the PyTorch
\cite{Paszke_NeurIPS2019}
environment.
We used ESPnet
\cite{Watanabe_IS2018}
for ASR model training and decoding.
We also used 
Hugging Face Transformers
\cite{Wolf_EMNLP2020}
with
the PEFT library
\cite{PEFT}
for LM
training,
domain adaptation,
and inference.
%
\subsection{Experimental settings}
\label{ssec_sets}
%
The CHiME-7 DASR task dataset
\cite{Cornell_CHiME2023}
consists of the three datasets,
i.e.,
CHiME-6
\cite{Watanabe_CHiME2020},
DiPCo
\cite{Segbroeck_arXiv2019},
and
Mixer 6
\cite{Brandschain_LREC2010}.
The former two datasets contain conversations
between four participants
at real dinner parties,
while Mixer 6 contains conversations
between an interviewer and a subject.
CHiME-6 and Mixer 6 have
the
training,
development (dev),
and evaluation (eval) data splits,
while DiPCo has the dev and eval data splits.
We used the CHiME-6 and Mixer 6
(CH6+Mx6) combined
training dataset for LM domain adaptation,
the CHiME-6 dev dataset for hyperparameter tuning,
and all the dev and eval datasets for evaluation.
Table~\ref{tab_data} shows details of these datasets,
and further details can be found in
\cite{Cornell_CHiME2023,Watanabe_CHiME2020,Segbroeck_arXiv2019,Brandschain_LREC2010}.
%
As described
in Section~\ref{ssec_text},
we sorted all the sentences (utterances)
in these datasets
in
the conversational order
(not the speaker-conditioned order \cite{Ye_CHiME2023})
and added a period for each sentence
(but we did not perform any capitalization).

%
For domain adaptation of Llama2,
we attached LoRA adapters \cite{Hu_ICLR2022}
to all the query and value projection matrices
in the attention modules of Llama2
and fine-tuned them
with QLoRA \cite{Dettmers_NeurIPS2023} (Section~\ref{ssec_adapt})
using
the CH6+Mx6 training dataset
shown in Table~\ref{tab_data}.
The ratio of the number of trainable parameters against
that
of all parameters was 0.06\%.
We set the context length (number of tokens)
$L$
in Eq.~(\ref{eq_cco})
at 0, 16, 32, 64, 128, 256, 512, and 1024, respectively.
%
For each of these context lengths $L$,
we concatenated past $L$ tokens as the context
to all the sentences in the dataset
and performed
fine-tuning.
We performed one epoch
QLoRA
fine-tuning
using the AdamW optimizer
\cite{Loshchilov_ICLR2019}
by setting
the LoRA rank,
LoRA alpha scaling parameter,
LoRA dropout probability,
batch size,
and learning rate
at 8, 16, 0.05, 64, 1e-5, respectively.
As a result,
we obtained
eight domain-adapted Llama2 models.

Table~\ref{tab_lms} shows the
configuration
of Slama2-70B (Section~\ref{ssec_lms})
in comparison with
that of Llama2-7B \cite{Touvron_arXiv2023a}.
We trained Slama2
using
1.1G tokens of
the LibriSpeech
text
dataset
\cite{Panayotov_ICASSP2015}.
We concatenated
all the sentences (token sequences) in the dataset
to form
one
long token sequence and split it into
token sequences of length 2048,
which is the maximum positional embedding length of Slama2,
as shown in Table~\ref{tab_lms}.
We trained Slama2 from scratch using these
token sequences
and then performed domain adaptation of it.
%
For each of the eight context lengths $L$,
we applied the same text processing described above
to the CH6+Mx6 training dataset
and performed
fine-tuning
of Slama2
using the dataset.
We performed one epoch full parameter fine-tuning using
the AdamW optimizer by setting the batch size and learning rate
at 64 and 5e-6, respectively. As a result, we obtained
eight domain-adapted Slama2 models.

As the E2E ASR model,
we trained a 
competitive
model
based on a Conformer-encoder
\cite{Gulati_IS2020}
and a structured state space (S4) decoder
\cite{Miyazaki_ICASSP2023},
which is used in the third-place system
\cite{Kamo_CHiME2023}
of the CHiME-7 challenge.
Using this ASR model,
we performed ASR for all the dev and eval utterances
and generated 32-best ASR hypotheses for each of the utterances.
We did not use any LMs in ASR decoding.
As with the above-described text processing,
we sorted the ASR hypotheses in the conversational order
and added a period for each hypothesis.
%
%
Then,
using Llama2, the domain-adapted Slama2/Llama2
of the eight context lengths $L$ (17 models in total),
respectively,
we performed rescoring for the 32-best ASR hypotheses.
%
When using Llama2,
we set the language weight $\alpha$ and the reward $\gamma$
in Eq.~(\ref{eq_nbest}) at 0.4 and 0.5, respectively,
and when using Slama2,
we set them at 0.3 and 0.5, respectively.
We optimized these values using the CHiME-6 dev dataset.
We also performed token-based PPL evaluation for all the dev and eval
transcriptions (correct token sequences).
%
\begin{table}[t]
\caption{\ftblskip Details of the CHiME-7 DASR task dataset. The numbers of words and tokens are counted using the manual transcriptions (correct sentences). However, we can obtain almost the same numbers with ASR hypotheses. \# tokens per word $\simeq$ 1.5 for all the datasets. For example, in the case of the CHiME-6 dev dataset, the context length L $=$ 1024 tokens corresponds to about 76 utterances (1024 / 13.4 $\simeq$ 76).}
\label{tab_data}
\vspace{-2.50mm}
\centering
\begin{tabular}{lrrrr}
\toprule
& \multicolumn{2}{c}{CH6+Mx6}
& \multicolumn{2}{c}{CHiME-6} \\
                     & \multicolumn{2}{r}{Training} &   Dev &  Eval \\
\midrule
\# utts (\# sents)   & \multicolumn{2}{r}{120k}     &  6.6k & 18.2k \\
\# words             & \multicolumn{2}{r}{994k}     & 58.9k &  101k \\
\# tokens            & \multicolumn{2}{r}{1.48M}    & 89.1k &  164k \\
\# words per utt     & \multicolumn{2}{r}{8.3}      &   8.9 &   5.5 \\
\# tokens per utt    & \multicolumn{2}{r}{12.4}     &  13.4 &   9.0 \\
\toprule
& \multicolumn{2}{c}{DiPCo}
& \multicolumn{2}{c}{Mixer 6} \\
                     &   Dev &  Eval &   Dev &  Eval \\
\midrule
\# utts (\# sents)   &  3.7k &  3.4k & 14.8k &  5.1k \\
\# words             & 30.0k & 28.8k &  149k & 69.3k \\
\# tokens            & 45.9k & 43.2k &  215k & 96.1k \\
\# words per utt     &   8.2 &   8.5 &  10.1 &  13.6 \\
\# tokens per utt    &  12.5 &  12.7 &  14.5 &  18.8 \\
\bottomrule
\end{tabular}
\end{table}
\begin{table}[t]
\caption{\ftblskip Configurations of Llama2-7B and Slama2-70M.}
\label{tab_lms}
\vspace{-2.50mm}
\centering
\begin{tabular}{lrr}
\toprule
                          & Llama2-7B & Slama2-70M \\
\midrule
Number of hidden layers   &        32 &          8 \\
Hidden size               &      4096 &        512 \\
Number of attention heads &        32 &          8 \\
Intermediate (FFN) size   &     11008 &       2048 \\
Max positional embeddings &      4096 &       2048 \\
\bottomrule
\end{tabular}
\end{table}
%
\subsection{Results of PPL evaluation and N-best rescoring}
\label{ssec_wers}
%
Table~\ref{tab_wers} shows the results of PPL evaluation
and $N$-best rescoring.
First,
we can confirm that,
in some cases,
the domain-adapted Slama2 reduces
the word error rates (WERs) from the strong ASR 1-best baseline.
The longer contexts bring the lower WERs (and PPLs).
%
However,
the reduction of the WERs is limited,
as reported
in the CHiME-7 papers
\cite{Ye_CHiME2023,Kamo_CHiME2023,Prisyach_CHiME2023,Park_CHiME2023}.

Next,
we compare the results of Slama2 and Llama2 without domain adaptation.
We can confirm that,
with the shorter context lengths
(especially when $L{=}0$),
Llama2 underperforms Slama2.
However,
its performance is quickly improved by
considering longer contexts,
i.e., by capturing the flow of a conversation.
%
It achieves the lowest WERs
by using a long context length, e.g., 512 and 1024.

Finally,
we compare the results of Llama2 and the domain-adapted Llama2.
We can confirm that,
unfortunately, domain adaptation does not bring further WER reduction.
However, it shortens the context length needed with Llama2 to achieve
the lowest WERs.
This is a large advantage since the computational cost of an LLM heavily
depends on the length of an input token sequence,
and by using shorter context lengths,
we can greatly reduce the computational cost.
For example,
the
inference
time when $L{=}128$ is about 1/10 of that when $L{=}1024$.
%
%
As reported in \cite{Udagawa_IS2022,Chen_ICASSP2023},
we also
confirmed
that recognition errors of infrequent words,
such as
``claustrophobic"
and 
``octogenarians",
were reduced by using Llama2.
Llama2 steadily reduces WERs from the strong ASR 1-best baseline,
but there is still room for
improvement since the
lowest WERs obtained with Llama2 are much
higher than those of the oracle hypotheses
shown in the last row of Table~\ref{tab_wers}.
%
\begin{table*}[t]
\caption{\ftblskip PPLs and N-best rescoring results in WERs obtained respectively with Llama2 and the domain-adapted Slama2/Llama2 of the eight context lengths $L$ (17 models in total) on the CHiME-7 DASR task dataset. WERs lower than the baseline ASR 1-best WERs are underlined, and the lowest WERs for each dataset are shown in bold font. If the WER reduction from the ASR 1-best WER is statistically significant at the 5\% / 1\% level, the WER is annotated with ``$\ast$" / ``${\ast}{\ast}$" \cite{Nakagawa_JASJ1994}. DiPCo is not included in the domain adaptation dataset (Table~\ref{tab_data}). Thus, the WER reductions on the DiPCo datasets are smaller than those on the CHiME-6 and Mixer 6 datasets.}
\label{tab_wers}
\vspace{-3.00mm}
\centering
\begin{tabular}{rrrllllllllllll}
\toprule
&
& 
& \multicolumn{4}{c}{CHiME-6}
& \multicolumn{4}{c}{DiPCo}
& \multicolumn{4}{c}{Mixer 6}
\\
&
& 
& \multicolumn{2}{c}{\hspace{-2.9mm}Dev} 
& \multicolumn{2}{c}{\hspace{-3.5mm}Eval}
& \multicolumn{2}{c}{\hspace{-0.3mm}Dev} 
& \multicolumn{2}{c}{\hspace{-0.6mm}Eval}
& \multicolumn{2}{c}{\hspace{-1.2mm}Dev} 
& \multicolumn{2}{c}{\hspace{-0.5mm}Eval}
\\
Model
& \makebox[7.0mm]{Adapt}   
& \makebox[2.0mm]{$L$}     
& \makebox[5.0mm]{PPL} & \makebox[5.0mm]{WER}
& \makebox[5.0mm]{PPL} & \makebox[5.0mm]{WER}
& \makebox[5.0mm]{PPL} & \makebox[5.0mm]{WER}
& \makebox[5.0mm]{PPL} & \makebox[5.0mm]{WER}
& \makebox[5.0mm]{PPL} & \makebox[5.0mm]{WER}
& \makebox[5.0mm]{PPL} & \makebox[5.0mm]{WER}
\\
\midrule
ASR 1-best 
& ---
& ---
& --- & 23.0 & --- & 26.2
& --- & 27.7 & --- & 25.5
& --- & 13.8 & --- & 15.8
\\
\midrule
\hspace{-1.2mm}Slama2-70M
& Full
& 0
& 48.3 & \underline{22.8}
& 48.3 & 26.2
& 48.4 & 27.8
& 45.6 & 25.8
& 46.3 & 14.0
& 45.4 & 15.9
\\
& 
& 16
& 44.4 & \underline{22.8}
& 41.2 & \underline{26.1}
& 44.6 & 27.7
& 41.3 & 25.7
& 41.8 & 14.0
& 42.1 & 15.8 
\\
& 
& 32
& 41.9 & \underline{22.8}
& 38.3 & \underline{26.0}
& 42.7 & 27.7
& 39.4 & 25.6
& 39.9 & 14.0
& 40.3 & 15.8 
\\
& 
& 64
& 39.5 & \underline{22.8}
& 36.0 & \underline{26.0}
& 40.9 & 27.7
& 37.3 & 25.6
& 37.9 & 14.0
& 38.3 & 15.8 
\\
& 
& 128
& 37.6 & \underline{22.8}
& 34.2 & \underline{26.0}
& 39.5 & 27.7
& 35.7 & 25.6
& 36.2 & 13.9
& 36.6 & 15.8 
\\
& 
& 256
& 36.4 & \underline{22.8}
& 32.9 & \underline{26.0}
& 38.5 & 27.7
& 34.6 & 25.6
& 35.1 & 13.9
& 35.4 & \underline{15.7}
\\
& 
& 512
& 35.7 & \underline{22.7}
& 32.2 & \underline{25.9}
& 38.0 & \underline{27.6}
& 34.1 & 25.6
& 34.4 & 13.9
& 34.7 & \underline{15.7}
\\
& 
& 1024
& 35.5 & \underline{22.7}
& 32.0 & \underline{25.9}
& 37.9 & 27.7
& 34.0 & 25.5
& 34.2 & 13.9
& 34.4 & 15.8 
\\
\midrule
Llama2-7B
& ---
& 0
& 57.6 & \underline{22.9}
&102.0 & \underline{26.1}
& 66.2 & 28.3
& 57.2 & 26.5
& 52.3 & 14.5
& 38.6 & 16.2 
\\
&
& 16
& 29.5 & \underline{22.6}
& 32.6 & \underline{25.7}${}^{\ast}$
& 32.6 & 27.9 & 27.5 & 26.0
& 25.3 & 14.2 & 22.5 & 15.9 
\\
&
& 32
& 22.9 & \underline{22.5}${}^{\ast}$
& 22.9 & \underline{25.7}${}^{\ast}$
& 25.0 & 27.8 & 21.5 & 26.0
& 19.0 & 14.1 & 18.0 & 15.8 
\\
&
& 64
& 19.0 & \underline{22.5}${}^{\ast}$
& 18.8 & \underline{25.5}${}^{\ast\ast}$
& 20.4 & 27.8 & 17.3 & 25.8
& 15.4 & 13.9 & 15.0 & \underline{15.7}
\\
&
& 128
& 16.8 & \underline{22.4}${}^{\ast}$
& 16.5 & \underline{25.4}${}^{\ast\ast}$
& 17.8 & \underline{27.5}
& 15.0 & 25.6
& 13.5 & \underline{13.7}
& 13.2 & \underline{15.6} 
\\
&
& 256
& 15.4 & \underline{22.3}${}^{\ast\ast}$
& 15.1 & \underline{25.4}${}^{\ast\ast}$
& 16.3 & \underline{27.5}
& 13.8 & \underline{25.5}
& 12.5 & \underline{13.6}
& 12.1 & \underline{15.5}
\\
&
& 512
& 14.6 & \underline{{\bf 22.2}}${}^{\ast\ast}$
& 14.1 & \underline{25.3}${}^{\ast\ast}$
& 15.5 & \underline{{\bf 27.3}}
& 13.1 & \underline{25.4}
& 11.9 & \underline{13.6}
& 11.4 & \underline{15.5}
\\
&
& 1024
& 14.1 & \underline{{\bf 22.2}}${}^{\ast\ast}$
& 13.5 & \underline{25.3}${}^{\ast\ast}$
& 15.0 & \underline{{\bf 27.3}}
& 12.7 & \underline{{\bf 25.3}}
& 11.6 & \underline{{\bf 13.5}}${}^{\ast}$
& 11.1 & \underline{{\bf 15.4}}${}^{\ast}$\hspace{-10.0mm}
\\
\cmidrule{2-15}
& \hspace{-1.2mm}QLoRA
& 0
& 20.9 & \underline{22.4}${}^{\ast}$
& 25.4 & \underline{25.7}${}^{\ast}$
& 23.5 & \underline{27.6}
& 20.4 & 25.7
& 19.5 & \underline{13.7}
& 16.9 & \underline{15.5}
\\
& 
& 16
& 18.4 & \underline{22.3}${}^{\ast\ast}$
& 18.0 & \underline{25.4}${}^{\ast\ast}$
& 20.2 & \underline{27.5}
& 17.2 & 25.6
& 15.2 & \underline{13.6}
& 14.3 & \underline{15.5}
\\
& 
& 32
& 16.8 & \underline{{\bf 22.2}}${}^{\ast\ast}$
& 15.9 & \underline{25.3}${}^{\ast\ast}$
& 18.5 & \underline{27.4}
& 15.6 & 25.5
& 13.8 & \underline{13.6}
& 13.1 & \underline{15.5}
\\
& 
& 64
& 15.5 & \underline{{\bf 22.2}}${}^{\ast\ast}$
& 14.7 & \underline{25.3}${}^{\ast\ast}$
& 17.2 & \underline{27.4}
& 14.4 & \underline{25.4}
& 12.7 & \underline{13.6}
& 12.3 & \underline{{\bf 15.4}}${}^{\ast}$\hspace{-10.0mm}
\\
& 
& 128
& 14.6 & \underline{{\bf 22.2}}${}^{\ast\ast}$
& 13.9 & \underline{25.3}${}^{\ast\ast}$
& 16.1 & \underline{27.4}
& 13.5 & \underline{{\bf 25.3}}
& 12.0 & \underline{{\bf 13.5}}${}^{\ast}$
& 11.7 & \underline{{\bf 15.4}}${}^{\ast}$\hspace{-10.0mm}
\\
& 
& 256
& 14.1 & \underline{{\bf 22.2}}${}^{\ast\ast}$
& 13.3 & \underline{{\bf 25.2}}${}^{\ast\ast}$
& 15.4 & \underline{27.4}
& 13.0 & \underline{{\bf 25.3}}
& 11.6 & \underline{{\bf 13.5}}${}^{\ast}$
& 11.3 & \underline{{\bf 15.4}}${}^{\ast}$\hspace{-10.0mm}
\\
& 
& 512
& 13.6 & \underline{{\bf 22.2}}${}^{\ast\ast}$
& 12.9 & \underline{{\bf 25.2}}${}^{\ast\ast}$
& 15.0 & \underline{{\bf 27.3}}
& 12.6 & \underline{{\bf 25.3}}
& 11.4 & \underline{{\bf 13.5}}${}^{\ast}$
& 11.0 & \underline{{\bf 15.4}}${}^{\ast}$\hspace{-10.0mm}
\\
& 
& 1024
& 13.4 & \underline{{\bf 22.2}}${}^{\ast\ast}$
& 12.6 & \underline{{\bf 25.2}}${}^{\ast\ast}$
& 14.7 & \underline{{\bf 27.3}}
& 12.4 & \underline{{\bf 25.3}}
& 11.3 & \underline{{\bf 13.5}}${}^{\ast}$
& 10.8 & \underline{{\bf 15.4}}${}^{\ast}$\hspace{-10.0mm}
\\
\midrule
Oracle
& ---
& ---
&   --- & 16.6
&   --- & 17.2
&   --- & 19.3
&   --- & 18.0
&   --- &  8.8
&   --- & 11.6
\\
\bottomrule
\end{tabular}
\vspace{-4.85mm}
\end{table*}
%
\subsection{Comparison of experimental settings}
\label{ssec_ppls}
\vspace{-1.00mm}
%
As described in Sections~\ref{ssec_lms} and \ref{ssec_text},
we performed PPL evaluation on the CHiME-6 dev dataset
to compare experimental settings
with the following three aspects,
i.e.,
(1) capitalize the first character of each sentence or not,
(2) sort utterances
in the conversational order or in the speaker-conditioned order
\cite{Ye_CHiME2023},
and
(3) use Llama2-7B or Llama2-7B-Chat \cite{Touvron_arXiv2023a}.

Table~\ref{tab_ppls} shows the experimental results.
The
leftmost setting is our current experimental setting described
in Section~\ref{ssec_sets}.
First,
we can confirm that,
by capitalizing the first character of each sentence,
the PPLs slightly get higher.
This result indicates
that capitalization is unnecessary
thanks to
the robust text processing ability of Llama2,
or we need a more sophisticated approach
for recovering the original text forms.

Next,
we can confirm that,
with the shorter context lengths,
the speaker-conditioned order shows the lower PPLs than those of the
conversational order,
but with the longer context lengths,
the trend is reversed.
This result indicates that
several consecutive utterances from one speaker have some consistency,
while, in the longer contexts,
the flow of a conversation becomes more dominant.

Finally,
we can confirm that,
by using Llama2-Chat,
the PPLs get
much higher.
This result indicates that
the
style
of the dialogue text datasets used to train Llama2-Chat
may be very different from
that
of the CHiME-7 DASR task dataset.
To summarize,
our current setting described in Section 4.1 seems to be reasonable.
%
\begin{table}[t]
\vspace{1.00mm}
\caption{\ftblskip Comparison results of the four experimental settings on the CHiME-6 dev dataset.}
\label{tab_ppls}
\vspace{-3.00mm}
\centering
\begin{tabular}{rrrrr}
\toprule
Llama2 version          &      7B &    7B &         7B & 7B-Chat \\
Utterance order         &    Conv &  Conv &       Spkr &    Conv \\
Capitalize the 1st char &      No &   Yes &         No &      No \\
\midrule
Context length $L$ $=$ 0 & {\bf 57.6} &  69.1 & {\bf 57.6} &    86.6 \\
                  16 &      29.5  &  31.6 & {\bf 28.4} &    41.2 \\
                  32 &      22.9  &  24.0 & {\bf 22.7} &    32.3 \\
                  64 & {\bf 19.0} &  19.9 &      19.4  &    26.3 \\
                 128 & {\bf 16.8} &  17.7 &      17.6  &    22.4 \\
                 256 & {\bf 15.4} &  16.2 &      16.4  &    20.0 \\
                 512 & {\bf 14.6} &  15.0 &      15.6  &    18.5 \\
                1024 & {\bf 14.1} &  14.4 &      15.1  &    17.7 \\
\bottomrule
\end{tabular}
\vspace{-5.00mm}
\end{table}
%
%
\section{Conclusion and future work}
\vspace{-1.00mm}
%
We investigated the applicability of LLMs for rescoring ASR hypotheses of 
highly casual conversations by using Llama2 \cite{Touvron_arXiv2023a} and
the CHiME-7 DASR task dataset \cite{Cornell_CHiME2023}.
%
Llama2 steadily
reduces WERs from
the strong ASR 1-best baseline
mainly with the effect of context-carry over.
Domain adaptation reduces the computational cost of Llama2
by shortening the needed context length.
%
The
experimental results
and findings
obtained in this study are informative
for researchers in this field.
Future work will include
using larger Llama2, i.e., 13B and 70B
\cite{Touvron_arXiv2023a},
and backward LMs
\cite{Xiong_IEEEACMTASLP2017,Irie_IS2018,Ogawa_ICASSP2022}.
%
%
\bibliographystyle{IEEEtran_ogawa}
\bibliography{ogawa}

\end{document}